\documentclass[a4paper]{article}
\usepackage{color}
\usepackage{RRA4}
\usepackage{hyperref}
\usepackage{xspace}

\newtheorem{ip}{Linear Program}
\newcommand{\bip}{\begin{ip}}
\newcommand{\eip}{\end{ip}}
\newcommand{\lp}{{\sc lp} \xspace}
\newcommand{\beas}{\begin{eqnarray*}}
\newcommand{\eeas}{\end{eqnarray*}}

\newcommand{\bw}{{\sc bw}\xspace}
\newcommand{\ieee}{\textsc{ieee} 802.11\xspace}
\newcommand{\somom}{\textsc{som}o\textsc{m}\xspace}

\RRtitle{Sur la capacité des réseaux ad hoc et hybrides à plat et auto-organisés}
\RRetitle{About the Capacity of Flat and Self-Organized Ad Hoc and Hybrid Networks}

\RRauthor{
Hervé Rivano \thanks[i3s]{I3S, UNSA -- INRIA MASCOTTE -- Email: herve.rivano@sophia.inria.fr}
       \and
Fabrice Theoleyre \thanks[citi]{Laboratoire CITI, INSA Lyon -- INRIA ARES -- Email: \{firstname.lastname@insa-lyon.fr\} }
       \and
Fabrice Valois \thanksref{citi} }

\RRnote{This work is partly supported by the European Commission, project IST-15964. The views given herein represent those of the authors and may not necessarily be representative of the views of the project consortium as a whole.}
\RRtheme{\THCom}
\RRprojet{MASCOTTE -- AR\`ES}

\URSophia
\URRhoneAlpes

\RRdate{September 2006}

\RRresume{ 
Les réseaux ad hoc concentrent un effort de recherche important sur la conception de protocoles efficaces. Les protocoles de routage basés sur une auto-organisation ont été principalement étudiés pour leurs propriétés de robustesse et de passage à l'échelle. Cependant, une auto-organisation pourrait peut-être engendrer une diminution de  la capacité réseau puisqu'elle concentre le trafic sur quelques liens radio privilégiés. Cet article présente quatre modèles pour évaluer la capacité d'un protocole de routage dans des réseaux \ieee. Notre approche consiste à modéliser comme des contraintes linéaires le partage de ressources radio de protocoles MAC tels que \ieee. Nous avons implémenté deux modèles d'équité. Le premier suppose que les n\oe uds ont un accès équitable au canal radio, tandis que le second suppose une équité entre liens radio. Ensuite, nous avons développé des scenarii de réutilisation spatiale optimiste et pessimiste du médium radio, créant respectivement une borne inférieure et supérieure de la capacité. Nos modèles sont indépendants du routage, et fournissent ainsi un framework pertinent pour leur comparaison. Nous avons appliqué ces modèles à une analyse comparative du protocole de routage du plus court chemin (OLSR) et de deux approches auto-organisées (VSR et Wu \& Li). Cette étude conclue sur la pertinence des approches auto-organisées du point de vue de la capacité réseau.
}

\RRabstract{
Ad hoc networking specific challenges foster a strong research effort on efficient protocols design. Routing protocols based on a self-organized structure have been studied principally for the robustness and the scalability they provide. On the other hand, self-organization schemes may decrease the network capacity since they concentrate the traffic on privileged links. This paper presents four models for evaluating the capacity of a routing schemes on 802.11 like networks. Our approach consists in modeling the radio resource sharing principles of 802.11 like MAC protocols as a set of linear constraints. We have implemented two models of fairness. The first one assumes that nodes have a fair access to the channel, while the second one assumes that on the radio links. We then develop a pessimistic and an optimistic scenarii of spatial re-utilization of the medium, yielding a lower bound and an upper bound on the network capacity for each fairness case.  Our models are independent of the routing protocols and provide therefore a relevant framework for their comparison. We apply our models to a comparative analysis of the well-known shortest path base flat routing protocol OLSR against two main self-organized structure approaches, VSR, and Wu \& Li's protocols. This study concludes on the relevance of self-organized approaches from the network capacity point of view.
}

\RRmotcle{capacité, auto-organisation, interférences radio, réseaux ad hoc, réseaux hybrides}
\RRkeyword{capacity, self-organization, radio interferences, ad hoc networks, hybrid networks}

\begin{document}
\makeRR

\section{Introduction}
MANet (Mobile Ad hoc NETworks) are spontaneous topologies of mobile nodes where each of them collaborate in order to propose services like routing, localization, etc. Each terminal can communicate via wireless links without preconditioned fixed infrastructure \cite{perkins94}. Moreover, the network must function autonomously, without any human intervention. To send packets from a source to a destination, either the destination is in the neighborhood of the source or intermediary nodes must relay the packets through a dynamic route. To reach this goal, the nodes must collaborate and exchange control information to set up routes in the network. Indeed, each node is both client and router. Because of the nodes mobility, radio links are created and deleted continuously leading to topology changes. So, self-adaptation of the network to the dynamicity is a major issue of MANet. Ad hoc networks thanks to their flexibility are promised to a large spectrum of utilization. They could be useful for military operations, allowing radio connections from vehicles to soldiers without systematical satellite communications. MANet could be used for rescue operations after a earthquake having destroyed all telecommunications infrastructures. More generally, MANet could be deployed in any scenario of spontaneous information sharing (conference, classroom, home,\dots).

Ad hoc networks can be connected to the Internet, via a dedicated device, the wireless access point (AP), constituting a gateway between the wired world and the ad hoc network. These networks are often called \emph{hybrid networks} constituting \emph{wireless multihops cellular networks}. We think that hybrid networks constitute a natural evolution of access networks. These networks could be used by telecommunications operator to extend cheaply the radio covering area of their cellular networks. Hybrid networks could also be intensively used in house automation networks, interconnecting personal services gateways at home to the Internet and their services providers.

Ad hoc and hybrid networks remain a large scientific domain to study. Classical networking solutions must be re-conceived because of the particular constraints of ad hoc networks: radio links implicate a low bandwidth, radio interferences, links instability creating rapid topology changes, a low reliability and packet losses. Moreover, ad hoc networks are mainly constituted by embedded terminals, presenting constraints in power-energy, CPU, memory\dots The network must collaborate to find a suitable power-energy saving policy. Several problems remain to be solved: addresses attribution, a solution to secure communications, an efficient interconnection to the Internet, a mobility management protocol, a routing protocol presenting high performances with an acceptable overhead\dots Finally, the flooding in ad hoc networks presents important problems of reliability, transmissions redundancy and collisions: the \emph{broadcast storm} problem \cite{ni99}.

In our point of view, self-organization can answer to the above key problems. The self-organization deals with virtual topologies in order to simplify ad hoc topologies. For example, virtual topologies can be constituted by a backbone \cite{wu03b}, or a combination of a backbone and clusters \cite{theoleyre04wcnc}. The goal is to offer control on the MANet. Advantages of virtual topologies are:
\begin{itemize}
	\item To take into account heterogeneous nodes, a virtual topology can classify {\em strongest} nodes as dominants and others as dominatees. Dominants  participate in the virtual topology in order to minimize the energy consumption (e.g.), hierarchizing the contributions.

	\item to hide neighborhood changes using a top-level view of MANet. A virtual topology stabilizes the neighborhood and simplifies the network topology

	\item to facilitate the integration of MANet in wireless/wired networks using the root of the virtual backbone. It can be viewed as a spontaneous wireless extension of wired infrastructures. It offers a natural way to manage hybrid networks

	\item to improve the scalability: because MANet are constituted by many nodes, clusters could group mobile nodes and a backbone could concentrate flooding packets to minimize the broadcast storm problem. So it is possible to provide scalable routing protocols: a local routing protocol restricted to the cluster only and a global routing protocol using the backbone
		
	\item to offer a framework to implement new services like mobility management, paging and localization services, native multicast support, etc
\end{itemize}
 
Because the self-organization seems to be an interesting way to manage MANet and because some privileged links and nodes are selected, the capacity of self-organized networks should be considered. When routing protocols based on self-organization are studied \cite{theoleyre05icc}, routes are often more robust but present a sub-optimal length. Moreover, because some nodes and links are privileged, a self-organization not well-conceived or exploited could create bottlenecks in the network and, finally, the capacity would decrease. Consequently, it would be interesting to study the impact of a self-organization on the network capacity and how the capacity is modified according to a virtual topology.

We focus here on the evaluation of the capacity of ad hoc and hybrid networks, i.e. the sum of the maximum throughput of all flow in the network. Through this evaluation, we propose to compare the capacity of self-organized and flat networks. This article makes two main contributions to the understanding of ad hoc networks. First, it provides a complete modeling of radio interferences using local nodes interactions, and gives lower and upper bounds on the achievable capacity, creating a distinction for the upper bound between the broadcast and the unicast traffic. We chose to model a fair medium access: each node has the same probability as its neighbors to transmit one packet. Secondly, it provides two definitions of the network capacity and compare flat and self-organized routing schemes.

Next, we will expose the related work about the evaluation of the capacity. Besides, we introduce classical and self-organized routing solutions. The section \ref{section:model} provides a full model of radio interferences and nodes interactions. Lower and upper bounds of the local capacity are also introduced. The section \ref{section:example} presents the study of two particular cases: the line and the grid. In the section \ref{section:evaluation} are proposed two definitions of the network capacity with two associated evaluation functions. Results are given in section \ref{section:results} in order to compare capacity of flat and self-organized routing protocols. Finally, we conclude the article and expose some perspectives.
\section{Related Work}
\label{section:related_work}

%
%

\subsection{Capacity Estimation}
Ad hoc networks seem promised to a large spectrum of applications. However, because of their structure, a user will not have the same throughput as in cellular networks. Hence, the capacity evaluation represents a key point to conceive adequate applications.


\subsubsection{The capacity of ad hoc networks}

Gupta \& Kumar \cite{gupta00} present a pioneering work in this domain. The authors propose an interference model based on the receiver. Let $range(u)$ be the radio range of the node $u$, and $d(u,v)$ be the distance from $u$ to $v$. The transmission from a node $u$ to a node $v$ is successful if $d(u,v) < range(u)$ and if there does not exist any other transmitting node $n$ nearer than $(1+\Delta) r(n)$ of $v$. 
The authors define the capacity as the aggregate achievable throughput. A throughput is achievable if each node in the network can send this quantity of packets according to spatial and scheduling constraints. 
The authors construct a Voronoi Tessellation, creating cells. Two cells can communicate directly if they have a common point. In the same way, two cells are interfering if there exist two points interfering with each other. According to the interference model, the number of interfering neighbors for a cell is in consequence bounded. The authors propose a scheduling scheme, and then a route construction approximating a shortest route in length. 
Let n be the number of nodes in the network. If the nodes are randomly placed on the sphere, the traffic is randomly distributed, and the radio range is identical for each node, the capacity per node is: 
\begin{equation}
	\Theta \left( \frac{1}{\sqrt{ n \cdot log(n)} } \right)
\end{equation}
Even when $m$ pure relay nodes are added uniquely to forward the traffic of the $n$ other nodes, the capacity is:
\begin{equation}
	\Theta \left( \frac{n+m}{n \sqrt{ (n+m) \cdot log(n+m)} } \right)
\end{equation}
However, if nodes are optimally placed, choosing optimally their radio range, the capacity per node is:
\begin{equation}
	O \left(  \frac{1}{\sqrt{n}} \right)
\end{equation}
In consequence, even in an optimal case, the capacity per node decreases when the number of nodes increases in the network. Intuitively, the increasing number of nodes allows an increasing spatial concurrency and frequency reuse of the radio medium. However, the route length increases in $O(\sqrt{n})$. The aggregate throughput of the  network increases in $O(\sqrt{n})$, giving finally the above result.

The authors present an asymptotic study of the capacity of ad hoc networks, independent of a particular routing protocol, or any distributed scheduling technology. This constitutes its strength and its weakness: this capacity does not reflect the radio medium sharing of 802.11 for example. Moreover, the interference model centered on the receiver is different from 802.11 for example: MAC acknowledgments being required, a communication is de facto (at least temporary) bidirectional. 
Finally, the capacity is inherently dependent from the routing protocols. Optimal routes will improve the global throughput, shortest routes can decrease it \cite{decouto03}. However, the authors don't take into account a not-optimal particular routing strategy. In consequence, although a study of the capacity proposed by different routing protocols could be interesting, such a work is not directly feasible with the proposition of Gupta \& Kumar.


\subsubsection{How to increase the capacity ?}

After this dramatic conclusion, \cite{grossglauser01} proposes to improve the capacity of ad hoc networks, taking into account the mobility. The goal is to decrease the route length: the less packet retransmissions occur, the more the capacity is improved. A source is a static node and forwards its packets to pure relay nodes. Relay nodes are mobile and forward the packet directly to the destination when it is directly in the radio range. The route length is in conclusion constant. As a consequence, the capacity per node remains constant in $O(1)$. However, the delay could be arbitrarily high, which is incompatible with the majority of applications. Moreover, the mobility of relay nodes is supposed completely random, which constitutes a strong hypothesis. \cite{bansal03b} tries to improve the capacity per node in limiting the end-to-end delay. However, some strong assumptions are done. First, the destination must be a static node. Secondly, the direction, speed and destination of the mobile nodes must be known by the node itself and its neighbors, requiring a GPS. Finally, the location of the destination must be known a priori, according to a scheme not presented in the article. In conclusion, we think such a scheme difficult to apply in realistic environments.


\subsubsection{The capacity of hybrid networks}

Several articles \cite{kozat03,liu03,zemlianov05} propose to extend the work of Gupta \& Kumar \cite{gupta00} to hybrid networks. \cite{kozat03,liu03,zemlianov05} use the same interference model as presented in \cite{gupta00}. Hybrid networks allow to deploy Access Points (AP) in the ad hoc network. These AP are connected via wired links: they constitute gateways to the Internet, or shortcuts in the network. Hence, these networks will surely be largely deployed in the future for domestic automation, for multihops Internet connectivity \cite{bruno05}\dots
It could be interesting to study the capacity evolution: it could be improved in using the wired backbone, or inversely reduced because of the creation of hotspots (AP could receive traffic from many clients and constitute a bottleneck).

\cite{hsieh01} presents a performance comparison of cellular and multihops networks. The authors propose to measure the network capacity in the following manner: a node chooses a random neighbor as destination. Thus, the traffic pattern creates a Maximal Independent Set. All other results in \cite{hsieh01} don't mention the traffic pattern, and moreover, the capacity is measured through simulations. Hence, the dependability of the simulator, parameters, environment could be problematic. The results are thus difficultly generalizable.

In \cite{kozat03}, nodes and Access Points (AP) are randomly distributed on a disk. The authors construct a Voronoi tessellation. Finally, if the ad hoc network is required to be connected without the help of the infrastructure, the capacity is:
\begin{equation}
	O\left(   \frac{1}{log(n)}  \right)
\end{equation}
If the network is connected, potentially through AP, the capacity is:
\begin{equation}
 	O\left(   \frac{1}{\sqrt{n log(n)}}    \right)
\end{equation}
This constitutes a little improvement in the capacity, although the capacity is not yet in $O(1)$.

In \cite{liu03}, $m$ AP are placed on a regular grid (consequently improving the capacity), and $n$ nodes are randomly distributed on a disk. A Voronoi tessellation is here also constructed. The radio range is uniform among the nodes. The routing strategy is as follows. The source sends directly the packet to the destination if it is in one of the $k$ nearest cells. Else, the packet is sent to the nearest AP. This AP will forward the packet to an AP in the cell of the destination. the parameter $k$ allows to propose a load balancing among the ad hoc and infrastructure nodes. The authors propose to study the global aggregated throughput. Hence, the traffic of particular nodes could be null. Such an unfairness is in contradiction with the definition of ad hoc networks. The impossibility to communicate for a node could be problematic for several applications. Let $T(m,n)$ be the global achievable traffic. The authors identify different scaling regime in the growth in the number of AP:
\begin{eqnarray}
	m =o(\sqrt{n}) 		  &	\Rightarrow &  T(m,n)=\Theta\left(
																												\sqrt{
																																	\frac{n}{    log\left(\frac{n}{m^2}\right)   }   
																															}
																										\right)\\
	m =\Omega(\sqrt{n}) &	\Rightarrow &  T(m,n)=\Theta(m)
\end{eqnarray}

In \cite{zemlianov05}, the authors propose to extend the work of \cite{liu03} in studying one additional scaling regime for the growth of the number of AP, in placing randomly both nodes and AP, introducing fairness among the nodes and in proposing an heterogeneous and adjusted radio range for AP. Let $\lambda(n,m)$ be the capacity per node with $n$ nodes and $m$ AP.

\begin{eqnarray}
	\label{eq:zemlianov_1}
	m \leq \sqrt{\frac{n}{log\ n}} 		                    & 		\Rightarrow		& 	\lambda(n,m)=\Theta\left(\frac{1}{\sqrt{n\ log\ n}}\right) \\
	\label{eq:zemlianov_2}
	\sqrt{\frac{n}{log\ n}} \leq m \leq \frac{n}{log\ n} 	& 		\Rightarrow		& 	\lambda(n,m)=\Theta\left(\frac{m}{n}\right)              \\
	\label{eq:zemlianov_3}
	m \leq n/log\ n        	             	                & 		\Rightarrow		& 	\lambda(n,m)=\Theta\left(\frac{1}{log\ n}\right)
\end{eqnarray}
The authors remark that in the first regime (eq. \ref{eq:zemlianov_1}), communications will uniquely use the ad hoc mode, the number of AP being too small. In the third regime (eq. \ref{eq:zemlianov_3}), the capacity reaches an asymptotic maximum even when the number of AP becomes arbitrarily high. This corroborates the results of \cite{kozat03}.


\subsubsection{Evaluation of the capacity through simulations}

In \cite{li01}, the throughput of the chain and grid topologies are studied. The authors show through simulations that 802.11 does not achieve the maximal theoretic capacity. Finally, they propose a traffic pattern which allows a scalable capacity: the problem being an increasing route length when the number of nodes increases, it is sufficient to limit this growth. The probability that a node communicates with another node at the distance $x$ is:
	\[
		p(x) = \frac {x^\alpha}  {\int^{\sqrt{A}}_{\epsilon} t^\alpha dt}
\]
With A being the surface area, $\epsilon$ the minimum distance from a source to a destination and $\alpha$ a parameter of the traffic pattern. The capacity is in $O(1)$ only if $\alpha < -2$. However, such a local traffic pattern can not be applied to all applications.

\cite{krause04b} studies the capacity of ad hoc networks through simulations. All nodes generate some traffic according to a rate $\mu$, with a random destination for each packet. A packet is forwarded along the shortest path in hops. The authors model the behavior of an ideal MAC layer: during the slot $t$, a random node $S_1$ with a not null packet buffer is firstly chosen. It sends its first packet in the buffer to the next hop $D_1$. As a consequence, $S_1$ and $D_1$ block all their neighbors, the interfering nodes. Thus, the interference model follows the transmitter/receiver model described above. Then, another node $S_i$ is randomly chosen. If it is not blocked, i.e. in interference with another node, it is selected and sends the first packet of its buffer to the next hop $D_i$. If $D_i$ is blocked, $S_i$ will try to send the first packet of its buffer which is intended to one node not blocked during the slot $t$. This allocation scheme seems for us near from the 802.11 behavior, as explained further. The capacity estimation is based on simulations without analytical study.


\subsubsection{AP placement to optimize the capacity}

\cite{qiu04} deals with the problem of capacity optimization in the placement of AP. The problem is to minimize the number of AP according to the quantity of traffic required by each node. The authors propose a LP formulation of the different constraints, and greedy algorithms allowing the placement of AP. They propose 3 interference models. However, the most sophisticated model proposes a throughput which decreases linearly with the route length, translating a macroscopic behavior, and deleting the local problems of radio resource sharing, important in radio transmissions.


\subsubsection{LP formulation of the problem of capacity}

In \cite{jain03}, \lp helps to model the capacity of ad hoc networks. The authors focus on the capacity when the topology and the traffic workload are given, which is the problem we treat. The LP formulation models the capacity problem as a multiflows problem. However, the complexity of such a formulation resides in the capacity estimation of each edge. Hence, the authors propose a lower and an upper bound. A conflict graph is constructed from the initial graph: a vertex is associated to each edge, and there exists one edge in the conflict graph between two vertices if the corresponding edges in the initial graph are interfering. 
The model of interferences proposed by the authors take into account the Signal to Noise Ratio. The transmission from $u$ to $v$ is successful if $SNR_{uv} > threshold$, the threshold being dependent of the network card. In the noise is taken into account the ambient noise, and the signal of other transmitters received in $v$.
In the lower bound, the authors estimate that the maximum throughput is achieved when a scheduling is contained in an independent set. A linear combination of schedulings of edges that lie in an independent set is also achievable. The authors propose to find several independent sets and to authorize at the time $t$ one single independent set. The capacity allocated to one edge $e$ is the sum of active independent sets in which it appears. In a fair MAC layer, each node receives a bandwidth proportional to the number of its neighbors trying to access to the medium. In  \cite{jain03} the authors propose to model an unfair MAC layer: all the MIS are considered equiprobable, although in a fair MAC layer, the probability of selection of a MIS depends on the nodes which constitute it and their order. We will see further that our approach takes into account such a distributed fair medium sharing, reflecting the desirable interactions of a distributed MAC layer.

\cite{kumar05} proposes a scheduling scheme maximizing the throughput from a group of sources to a group of destinations. The article deals with several interferences models:
\begin{itemize}
	\item Transmitter model: a node $u$ can communicate with a node $v$ if no node $w$ exists nearer than $(1+\Delta)\cdot (range(u) + range(w))$ from $u$. The radio range can be heterogeneous among the nodes
	\item Protocol model: the interference model of \cite{gupta00}
	\item Transmitter/receiver model: two edges can be activated simultaneously if they are more than 2 hops far.
\end{itemize}
The authors propose a scheduling of radio links so that no two links activated simultaneously are interfering. The authors propose an algorithm to allocate locally the capacity. A greedy algorithm allocates slots to the different edges according to their decreasing euclidean length. A condition is given so that the allocation is achievable: the capacity allocated to one edge $e$ and to all edges longer than $e$ and in interference with $e$ is inferior to the radio bandwidth. Such a scheduling achieves a throughput being at most $k$ less than the optimal throughput, $k$ being a constant. The authors tend to underestimate the quantity of traffic which could be sent: when 2 edges $e_1$ and $e_2$ are interfering with one edge $e$, the capacity is shared among $e$, $e_1$ and $e_2$. However, $e_1$ and $e_2$ could perhaps transmit packets simultaneously. Our capacity estimation proposes a finer evaluation of the local resource sharing in studying more precisely the interference interactions among the 2-neighborhood of a node. Finally, the authors propose an interesting metric to model the fairness in ad hoc networks: they limit the ratio of the minimal throughput and the maximal throughput in the network by a constant. This allows to limit the discrimination of some flows.

%
%

\subsection{Routing protocols}

Routing protocols are very closely related to the capacity of ad hoc networks. According to the selected routes, a network will not achieve the same throughput. If a protocol selects shortest routes in hops, the throughput will not be maximal. Oppositely, if a protocol discovers routes in order to distribute the load in the network, avoiding the formation of hotspots and bottlenecks, the global throughput will be improved. We present here a short panorama of classical routing protocols which are used in ad hoc networks.


\subsubsection{Flat Approaches}
Routing is one of the major issue in ad hoc networks: a good delivery ratio with a low delay and overhead must be achieved. Hence, many propositions were done \cite{dsr_draft,aodv_rfc,olsr_rfc}. Two major approaches were proposed: the reactive and the proactive protocols.

In the proactive approach, a node knows a priori a route toward each other node in the network. If the whole topology is known, it can compute optimal routes using the Dijkstra algorithm for example. Such a solution could be interesting if a node communicates with several destinations, changing frequently along the time. Moreover, the delay is reduced and optimal routes could be computed according to several criteria (hops, congestion,\dots). However, proactive protocols require the periodical flooding of topology packets. As we said, floodings cause quickly the broadcast storm problem: the medium is heavily loaded, many collisions occur\dots OLSR \cite{olsr_rfc} proposes a solution to optimize the flooding, the \emph{Multi Point Relays}: a node chooses a set of 1-neighbors covering entirely its 2-neighborhood. Only these selected 1-neighbors are authorized to forward the topology packets coming from this node. Nevertheless, the overhead remains important.

In the reactive approach, a node does not know initially any route. It tries to discover a new route only when needed, \emph{on demand}. In AODV \cite{aodv_rfc}, a node which wants to send a \texttt{data packet} and which does not have a route toward a destination sends a \texttt{Route Request}. This request is flooded in the network: each node which receives the packet forwards it in broadcast. Any node which knows a route to the destination (in the worst case the destination itself) can send a \texttt{Route Reply}, forwarded along the inverse route. Hence, the reactive protocols require a delay to set up the route, before sending any \texttt{data packet}. However, the overhead could be reduced if a node has only a few corresponding nodes. Flooding being here also used massively, the broadcast storm can occur in the same way. Some mechanisms must be proposed to limit the impact of such a flooding.

To model the behavior of a flat approach, we consider OLSR as a representative protocol. Since we use a simplified representation of a routing protocol, only some special characteristics are useful: the shortest routes computed by OLSR are extracted from the topology and the average overhead per node is measured and injected in our LP formulation, as described later.


\subsubsection{Self-organized approaches}
Self-Organization through the construction and the maintenance of a virtual structure is an important topic in ad hoc networks \cite{sinha99,theoleyre04wcnc,wu03b}. A virtual structure consists in constructing a virtual topology on the radio topology: each node controls the identity of its virtual neighbors, being a subset of its radio neighbors. In the same way, some links are not used whereas over are privileged.

\cite{wu03b} proposes to select some nodes to act as backbone members. The election process is completely local, creating a negligible overhead. This backbone could be used for the flooding optimization: a node sends a packet in broadcast, and only backbone members are allowed to forward this packet. Hence, the gain of such a structure for flooding is $\frac{n}{\left|BACKBONE\right|}$, $n$ being the number of nodes in the network and $BACKBONE$ being the set of backbone nodes. A node is declared dominatee, i.e. backbone client, if a connected set of neighbors with an higher weight constitutes a dominating set of its whole neighborhood. Else, it is dominator. The weight can be based on address, degree, energy\dots The cardinality of this CDS is clearly not the cardinality of the Minimum Connected Dominating Set of the graph. However, this proposition is a good trade-off between cardinality and overhead. \cite{wu01b} presents a routing protocol for such a structure: a node forwards its data packet to the nearest backbone member. Then, a classical routing protocol is executed on the backbone topology. However, the backbone could constitute perhaps a bottleneck in the network, and more importantly such a routing scheme is not particularly adapted to the virtual structure organization.

\cite{theoleyre04wcnc} presents a virtual structure composed by both a virtual backbone and clusters. The backbone allows to optimize information floodings and to create a distinction between the \emph{clients} and the backbone members, acting as \emph{masters}. Clusters group together nodes geographically close. These clusters represent flexible services areas, with one leader per area, the \emph{clusterhead}. In conclusion, such a structure self-organizes the network and simplifies further services deployment like routing, addresses attribution\dots Distributed procedures for construction and maintenance were proposed to adapt automatically the structure to the topology changes. Moreover, the algorithms are self-stabilizing \cite{theoleyre05sss}.

VSR \cite{theoleyre05icc} is a routing protocol for ad hoc networks adapted to such a virtual structure. It is based and designed to demonstrate that a self-organization is useful to route packets. A node knows routes in its cluster, using a proactive protocol inside the cluster. Oppositely, if a node must reach a destination outside its cluster, a route is discovered reactively using the virtual backbone only. A route is in this case constituted of a list of clusters id., stabilizing the route. The backbone is useful to optimize the flooding caused by the route discovering. Hence, some edges are not used for the flooding. We propose here to quantify the impact of this topology reduction.

\cite{theoleyre05pwc} presents a routing and localization protocol for hybrid networks. In hybrid networks, the AP constitutes the single destination: nodes want to communicate uniquely with hosts in the Internet. Such a routing scheme benefits from a virtual backbone. In upload, a proactive route is known a priori toward the root of the backbone, the AP. In download, the AP can discover reactively a route toward one of its clients, the virtual backbone helping to minimize the impact of flooding. Moreover, a gratuitous inverse route is created on the fly when a node initiates a connection to the Internet. Hence, the overhead of reactive route discovering will remain very low. Besides, as an extension, a routing scheme is proposed to route packets directly in ad hoc mode, without passing through the AP.

As a virtual topology is a subset of the radio topology maintaining the same radio range, it could be interesting to study the capacity associated to self-organization structures. To evaluate the impact of the self-organization on the capacity, we study VSR \cite{theoleyre05icc}, \textsc{som}o\textsc{m} \cite{theoleyre05pwc} and Wu \& Li \cite{wu01b} approaches. More precisely, only a few characteristics of the protocols are useful in our models: the routes actually computed and the average overhead associated to each node in the network.

\section{Hypothesis}
\label{section:model}

In this section we introduce two models of radio resource sharing for wireless {\sc ieee} $802.11$-like networks.

Unlike other approaches described in Section \ref{section:related_work}, our models aim at evaluating the capacity that is available to the nodes for an input couple (network,routing strategy).

More precisely, a discrete-event simulation process provides for the input data of our models: the network topology, the routes defined by the chosen routing protocol, and the rates of the control traffic which is locally broadcasted by each node to their neighborhoods.

Eventually, our approach is based on the following key statements:

\begin{itemize}
	\item The radio resource available to a node depends on the activity of its neighbors and on the MAC protocol. We focus on acknowledged protocols (and consequently to bidirectional links).

	\item These resource sharing processes induce localized sets of constraints on the bandwidth available for each node. The pessimistic and optimistic models mainly differ by these constraints.

	\item We consider the end to end traffic as network flows that induce load on the links they cross, hence network-wide global constraints. This global view of the network spares considering each data packet individually, resulting in an aggregated, high level view of the data transportation.

	\item Combining global and local constraints yields a linear model for the transport capacity of the whole network. Different notions of capacity might be defined, as detailed below, depending on the characteristics of the network that is studied.

	\item The linear constraints may exploit combinatorial and mathematical results on wireless networks. Most of these results are based on graph theory and stochastic analysis, and allow to save thousands of packet-level simulation iterations.
\end{itemize}

Our linear programing models fit the generic form described as \lp\ref{lp:generic}. Radio resource sharing constraints are defined for each node. The data flow load constraints added for each route define global constraints on the transport capacity of the network.

The set of the routes is given as ${\cal P}\ =\ \{p=(s,u_1,\dots,d)\}$, with $s$ being the source, $d$ the destination, and $u_i$ the $i^{th}$ intermediary node in the path. The objective function gives the capacity that we want to evaluate. Traffic management equations describe the flows that are to be carried by the network when data are sent on a path $p$.

\vbox{
\medskip
\bip[Generic model]~
\label{lp:generic}
\medskip
\begin{center}
\begin{tabular}{c c}
	\vspace{0.3cm}
		{\bf Maximize}																	&			Objective function on {\cal P} \\
	\vspace{0.2cm}
																										&			{\bf Subject to}\\	
		Resource sharing constraints 										&			$\forall$ node u   \\
	\vspace{0.2cm}
		around $u$																			\\
	\vspace{0.2cm}
		Traffic management for p												&			$\forall$ path p	
\end{tabular}
\end{center}
\eip
}

In the following, we first describe some hypothesis and notations.  Then, we present a set of resource sharing constraints which gives a pessimistic model of the MAC layer behavior, resulting in a lower bound on the overall capacity of the network. Finally, we give an optimistic model, hence resulting in an upper bound. This model is based on combinatorial and stochastic considerations on the MAC layer behavior.

\subsection{Common hypothesis and notations}
\label{sec:hypo}

In order to develop a linear model of the radio resource sharing, we need to assume some classic hypothesis on the MAC layer. Moreover, we assume a typical client/server traffic model, as follows:
\begin{enumerate}
	\item Perfect radio channel: we assume that the medium delivers a constant bandwidth and does not corrupt data transmissions.
	\item Ideal MAC layer: there are no collision, the bandwidth can be optimally used, and the probability to access to the medium is uniform. 
	\item Bi-directional unicast communications: if a node $u$ sends a data traffic to one of its neighbor $v$, $v$ answers with an acknowledgment: any other node interfering with $u$ or $v$ cannot access to the medium. 
	\item Control and topology maintenance traffic is sent by $u$ through a local broadcast preventing all nodes within 2 hops of $u$ (2-neighbors) from sending or receiving any kind of traffic.
\end{enumerate}

We are using the following notations:
\begin{itemize}
\item \bw is the available radio bandwidth. This gives the maximum amount of data that can be sent by one terminal.
\item $f(p)$ is the throughput of the data sent on the path $p$.
\item Let $u$ be a node. $T(u)$ is the total amount of traffic sent by
  $u$: $T(u) = \sum_{v\in\Gamma(u)}T(u,v) + T_{c}(u)$ with
  \begin{itemize}
	  \item $\Gamma_k(u)$ is the $k$-neighborhood of $u$, i.e. the set of the nodes at most $k$ hops far from $u$. $\Gamma_1(u)$ is written $\Gamma(u)$ for short. Note that $u\in\Gamma_k(u),\,\forall\,k$.
	  \item $\Delta_k(u)$ is the size of the $k$-neighborhood of $u$: $\Delta_k(u)\ =\ |\Gamma_k(u)|$.
	  \item $T(u,v)$ is the unicast traffic on the physical link $(u, v)$. 
	  \item $T_{c}(u)$ is the broadcast traffic for control and topology maintenance sent by $u$ to its neighborhood.
  \end{itemize}
\end{itemize}

Note that it is straightforward to write the traffic management constraints of \lp \ref{lp:generic} from the composition of $T(u,v)$.
\section{Models of radio resource sharing}
We propose two models of radio resource sharing: the node-oriented fairness, and the link-oriented fairness. In the first model, each vertex which tries to access to the radio medium receives the same capacity. This models well the classical MAC layer protocols. {\sc ieee} 802.11 introduces a random back-off delay for the nodes which want to access to the medium. In optimal conditions, all the nodes will have the same probability to have the medium. In the link-oriented fairness, each radio link in an interference area will receive the same quantity of capacity. According to us, this allows to avoid bottlenecks: some nodes have many links, and must forward an important quantity of traffic. If these nodes receive a capacity not proportional to the number of neighbors, they will constitute a bottleneck. 

For each model of fairness, we propose a pessimistic radio resource sharing constituting a lower bound of the capacity, and an optimistic radio resource sharing constituting an upper bound.

%
%

\subsection{Node-oriented fairness}
\label{section:vertex_fairness}
\subsubsection{A pessimistic resource sharing scenario}
\label{sec:vertices_lowerB}
A lower bounding linear program is obtained when local radio resource sharing constraints model a pessimistic behavior of the MAC layer. Indeed, two simultaneous communications (A$\rightarrow$B) and (C$\rightarrow$D) are achievable only if (A,C), (A,D), (B,C) and (B,D) are not neighbors. A 2-contention exists. 

\begin{figure}
	\centering
		\includegraphics[width=5cm]{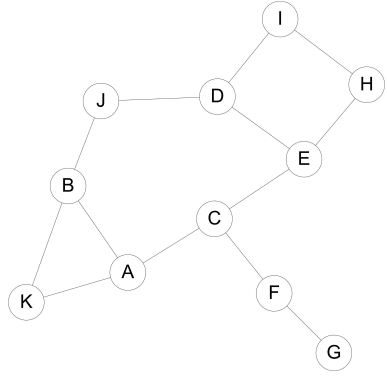}
	\caption{The 2-Neighborhood of one node}
	\label{fig:2_neighborhood}
\end{figure}

Thus, a pessimistic resource sharing is achieved if the transmission of one node is blocking all its 2-neighborhood (fig. \ref{fig:2_neighborhood}). If the center or one of its 2-neighbors transmits a data packet, no other node in the 2-neighborhood of the center is allowed to send packets. To model a node-oriented fairness, the same capacity is allocated to each 2-neighbors of the center. Naturally, this represents a worst case, since for example the couples (A,B)/(D,E)  or (K,B)/(H,I) can communicate simultaneously in the figure~\ref{fig:2_neighborhood}. 

One can notice that stopping any radio activity includes refusing any incoming connexion request, since it requires to send an acknowledgment for the received packet, creating collisions.

Applying the worst case of control traffic to data communications leads to a pessimistic MAC local behavior as follows. Let be a node $c$, called {\em the center} of its 2-neighborhood. 

\begin{itemize}
\item The radio bandwidth is uniformly distributed between all nodes in potential contention: the center $c$ and its whole 2-neighborhood, $\Gamma_2(c)$:
		\begin{eqnarray}
			\label{eq:vertex_low_bound_capa}
			 \forall c,\ \forall u \in \Gamma_2(c), \hspace{1cm}			 
			 T(u) \leq \frac{BW}{\Delta_2(c)}
		\end{eqnarray}
Note, that for each center $c$, a set of $\Delta_{2}(c)$ equations is given. In consequence, the capacity $T(u)$ allocated to a node $u$ is constrained by $\Delta_2(u)$ equations (one for each possible center).

	\item In the capacity allocated to one node, all the control traffic and the unicast transmissions must be scheduled. Additionally, a node allocates an equal capacity to each of its neighbors:
		\begin{eqnarray}
			\label{eq:vertex_low_bound_control}
			\forall u, \forall v \in \Gamma(u)-\{u\},\hspace{1cm}
			T(u,v) \leq \frac{T(u) - T_c(u)}{\Delta(u)-1}
		\end{eqnarray}
\end{itemize}
The equation set (\ref{eq:vertex_low_bound_control}) models the way each node manages its available bandwidth, while set (\ref{eq:vertex_low_bound_capa}) model the way the radio medium capacity is shared among the nodes.

Finally, we can remark that two nodes can send data simultaneously only if they are sufficiently distant, at least 3 hops, and a less loaded node is given the same bandwidth than a fully loaded one. This set of local constraints yields \hbox{\lp \ref{lp:vertex_lower_bound}}, whose solutions lower bound the total amount of traffic supported by the network.

\vbox{
\medskip
\bip[Pessimistic model]~
\label{lp:vertex_lower_bound}
\medskip
\begin{center}
\begin{tabular}{c c}
	\vspace{0.3cm}
		{\bf Maximize}																	&			Objective function on {\cal P} \\
	\vspace{0.2cm}
																										&			{\bf Subject to}\\	
	\vspace{0.2cm}
		Equation set (\ref{eq:vertex_low_bound_capa})   	& 		node c, the center	\\
	\vspace{0.2cm}
		Equation set (\ref{eq:vertex_low_bound_control})&			$\forall$ node $u$\\
	\vspace{0.2cm}
		Traffic management for p					&			$\forall$ path p	
\end{tabular}
\end{center}
\eip
}

\subsubsection{An optimistic resource sharing scenario}
\label{sec:vertices_upperB}
The pessimistic radio resource sharing model tends to over-estimate the interferences. Some communications could be possible in a realistic protocol, but are forbidden in our model. Hopefully, many protocols, like {\sc ieee} 802.11, achieve a better repartition. All the 2-neighborhood is not blocked. For example, in fig.\ref{fig:2_neighborhood}, the simultaneous transmissions (A$\rightarrow$B) and (D$\rightarrow$E) are possible, because the packets are not to be understood by $C$.

In consequence, we propose here an optimistic resource sharing model which allows several simultaneous transmissions in the 2-neighborhood of a node. This optimistic resource sharing is translated in local constraints. This constitutes an upper bound since we assume the existence of a global scheduling respecting all the local constraints. In other words, the local schedulings are assumed to be compatible with the global unit of all local schedulings.

Let assume that the canal is free. One neighbor, $u$, of the center sends a packet to $v$. All neighbors of $u$ will stop any activity. When $u$ finished the transmission, $v$ will send a MAC acknowledgment. Thus, no neighbor of $u$ or $v$ must be authorized to send a packet in order to avoid collisions. 

Let assume that another node $u'$ wants to send a packet. If $u'$ is neither neighbor of $u$ nor of $v$. $u'$ can send a packet. But it must choose a destination $v'$ which is not neighbor of $u$ or $v$. We model the medium as a central entity which allocates capacity to each edge, and avoiding interfering edges to transmit packets simultaneously. This hypothesis is strongly linked to the combinatorial concept of {\em independent set}.

Indeed, such a contention-free communication set is an independent set, maximal for inclusion, of the graph $L_{1,2}\left(\mathcal{L}\left(G_c\right)\right)$, defined as follows:

\begin{itemize}
	\item $G_c$ is the graph of the 2-neighborhood of $c$.
	\item ${\cal LG} = {\cal L}(G_c)$ is the {\em linegraph} of $G_c$,   that is the graph with one vertex per arc of $G_c$, and an edge between any two vertices  whose corresponding arcs are adjacent.
	\item $L_{1,2}({\cal LG})$ is the graph with the same vertices as $\cal LG$, and an edge between any two neighboring or 2-neighboring vertices (its 2-closure).
\end{itemize}

Independent vertices (i.e. pairwise non adjacent vertices) of $L_{1,2}\left(\mathcal{L}\left(G_c\right)\right)$ correspond to
contention-free communications. An inclusion-wise maximal independent set is therefore an inclusion-wise maximal set of communications that can be activated simultaneously.

Eventually, the MAC layer achieves a fair sharing of the bandwidth among the maximal independent sets. fairness of the nodes must be respected for this sharing. Let $BW(I)$ be the bandwidth given to the independent set $I \in {\cal I}$, the set of all maximal independent sets of $L_{1,2}\left(\mathcal{L}\left(G_c\right)\right)$. $BW(I)$ is proportional to $P(I)$, the probability of $I$ to be selected, and the bandwidth in the neighborhood of $c$ is shared as follows:

\begin{eqnarray*}
	\label{eq:upperB}
	BW(I) & =			& P(I) \cdot \left(BW - T(c) - \sum_{u \in \Gamma(c)-\{c\}} T_c(u)\right)\\
	\Rightarrow BW		&	\geq 	&	T(c) + \sum_{I\in{\cal I}}BW(I) + \sum_{u\in\Gamma(c)}T_c(u)
\end{eqnarray*}

The total bandwidth allocated to a communication link $(u,v)$ is the sum of the bandwidth allocated to each independent set including $(u,v)$:

\begin{eqnarray*}
	T(u,v) &	\leq	&	 \sum_{I \ni (u,v)} BW(I) \\
	T(u,v) &	\leq 	&	\left(BW - T(c) -\hspace{-0.3cm} \sum_{x \in \Gamma(c)-\{c\}} \hspace{-0.1cm} T_c(x)\right) \sum_{I \ni (u,v)} P(I)
\end{eqnarray*}

Moreover, $\sum_{I \ni (u,v)} P(I)$ is exactly the probability for the communication link $(u,v)$ to be activated by the canal. This quantity is hence denoted $P(u,v)$ in the following.

\begin{eqnarray}
	\forall (u,v) \in \Gamma_2(c)-\{c\}, \hspace{1cm}
	T(u,v) \leq \left(BW - T(c)  -\hspace{-0.3cm} \sum_{x \in \Gamma(c)-\{c\}} \hspace{-0.1cm} T_c(x)\right) P(u,v)
	\label{eq:UBfreq}
\end{eqnarray}

Unfortunately, on arbitrary network topologies, $P(I)$ and $P(u,v)$ cannot be computed unless the whole set $\cal I$ is known, and $\cal I$ has an exponential size. We therefore build a stochastic estimation of $P(u,v)$, denoted $freq(u,v)$ in the following. 

These frequencies $freq(u,v)$ must absolutely take into account the fairness among the nodes. We propose in consequence the following algorithm to construct an independent set:
\begin{itemize}
	\item While at least one not blocked node exists, take randomly one, say $u$
		\item chose randomly one neighbor $v$ of $u$ which is not blocked
			\begin{itemize}
				\item If $v$ exists, activate the communication $(u,v)$ and mark all the neighbors of $u$ and $v$ as blocked
				\item else, mark $u$ as blocked
			\end{itemize}
\end{itemize}
If this algorithm is repeated $n$ times, $freq(u,v)$ is equal to the proportion of the cases where the edge $(u,v)$ was selected. Note that each edge is directed: the edge $(u,v)$ will not receive the same amount of traffic as $(v,u)$.

In order to complete the model, control and topology maintenance traffic generated by the routing algorithms is to be completely taken into account. The equation set (\ref{eq:UBfreq}) models that when a neighbor of the center $c$ emits control traffic, $c$ has to stop any radio activity.  On the other hand, the $2$-neighbors of $c$ will include their control traffic into their allocated bandwidth.

The bandwidth is distributed per link: a node is not allowed to distribute locally the traffic to each of its link. Indeed, the capacity allocation take into account the specificities of each link, and particularly the fact that several links can send information simultaneously. If the node chooses to redistribute the capacity of an unloaded link to another of its links, the interference constraints could be violated. Such a behavior must be avoided.

The last optimistic aspect of this model is that the combinations of the local constraints might not yield a feasible share
of the global capacity. As a matter of fact, the union of the local independent sets might not be a global independent set. In other words, the global constraints are stronger that the union of the local ones. The linear program \ref{lp:vertex_upper_bound} neglects this fact, yielding an upper bound on the global capacity of the network.

\vbox{
\medskip
\bip[Optimistic model]~
\label{lp:vertex_upper_bound}
\medskip
\begin{center}
\begin{tabular}{c c}
	\vspace{0.3cm}
		{\bf Maximize}										&			Objective function on {\cal P} \\
	\vspace{0.2cm}
																			&			{\bf Subject to}\\	
	\vspace{0.2cm}
		Equation set (\ref{eq:UBfreq})		&			$\forall$ link $(u,v)$  $\in$ $E$\\
	\vspace{0.2cm}
		Traffic management for p					&			$\forall$ path p	
\end{tabular}
\end{center}
\eip
}

\subsubsection{Flexibility of the models}
We use the transmitter-receiver model for the representation of interferences \cite{kumar05}: two links can be activated simultaneously if they are at least 2 hops far in the linegraph $\mathcal{L}\left(G_c\right)$. However, our approach is generic for any other interference model. In the lower bound, we can construct the set containing one node $C$ and all the nodes which can interfere with its transmissions. The lower bound will distribute the same capacity to each node in this set.

In the same way, the upper bound could be constructed with any interference model. The linegraph $\mathcal{L}\left(G_c\right)$ will simply be replaced by any conflict graph (two edges are neighbors in a conflict graph if they are interfering). Two edges can be simultaneously activated if they are not neighbors in the conflict graph. Thus, in computing frequencies $freq(u,v)$, a node $n$ is considered as blocked if no edge $(n,x)$ exists such that $(n,x)$ and $(u,v)$ are not neighbors in the conflict graph.

In the previous models, we chose to present only the transmitter-receiver interference model for a sake of clarity for the explications.

%
%

\subsection{Link-oriented fairness}
Here are presented lower and upper bounds with a fairness oriented on edges. Instead of distributing an equal capacity to each node in competition to access to the medium, here is distributed much capacity to nodes which have more neighbors, and thus more traffic to forward. In consequence, is proposed here a link-oriented fairness. Since the previous models in section \ref{section:vertex_fairness} are very similar, we chose to present only the major differences.

Let introduce the following notation:
\begin{itemize}
	\item $\mathcal{LG}$: the linegraph of $\mathcal{G}$\footnote{$\mathcal{LG}$ is the linegraph of $G$, cf. previous paragraph for the exact definition}
	\item $\gamma_k$: the k-neighborhood in $\mathcal{LG}$ of one edge $e$ in $G$. Each link $e$ is directed.  
	\item $\delta_k$: $|\gamma_k|$ 
\end{itemize}

\subsubsection{A pessimistic resource sharing scenario}
\label{sec:edges_lowerB}

We keep on proposing a pessimistic radio resource sharing. However, instead of distributing the capacity to vertices, we construct for each edge in the graph the set of its neighborhood in the conflict graph. If one edge is active, it is potentially in conflict with each edge in this set. Moreover, the same capacity is allocated to each edge. 
In consequence:

\begin{eqnarray}
	 \forall e \in E, \forall f \in \gamma_2(e), \hspace{1cm}
	 T(f) \leq  \frac{BW - \sum_{(u,x) \in \gamma_2(e) } T_c(u) 	 }{\delta_2(e)}  
	 \label{eq:edge_low_bound_capa}
\end{eqnarray}

\medskip

Additionally to data packets, a node must send control traffic. The equation \ref{eq:vertex_low_bound_control} of the previous lower bound keeps on holding. 
Finally, we obtain the following linear program {\sc lp} \ref{lp:edge_lower_bound}:

\vbox{
\medskip
\bip[Pessimistic model]~
\label{lp:edge_lower_bound}
\medskip
\begin{center}
\begin{tabular}{c c}
	\vspace{0.3cm}
		{\bf Maximize}																			&			Objective function on {\cal P} \\
	\vspace{0.2cm}
																												&			{\bf Subject to}\\	
	\vspace{0.2cm}
		Equation set (\ref{eq:edge_low_bound_capa})  	& 		edge e	\\
	\vspace{0.2cm}
		Equation set (\ref{eq:vertex_low_bound_control})		&			$\forall$ node $u$\\
	\vspace{0.2cm}
		Traffic management for p														&			$\forall$ path p	
\end{tabular}
\end{center}
\eip
}

We can remark that the lower bounds with link-oriented and node-oriented fairness are not comparable. A different behavior of the MAC layer is modeled. Thus the capacity of the network depends on the protocol chosen to allow concurrent access to the medium.

\subsubsection{An optimistic resource sharing scenario}
\label{sec:edges_upperB}
The upper bound with an link-oriented fairness knows less modifications. The only detail to change is the amount of capacity to distribute to each edge. Indeed, only the algorithm computing the $freq(u,v)$ (capacity allocated to each link in the 2-neighborhood of one node) must be changed. We propose the following new algorithm:

\begin{itemize}
	\item While at least one not blocked edge exists, take randomly one, say $(u,v)$
	\item Mark as blocked any other edge interfering with $(u,v)$
\end{itemize}
If this algorithm is repeated $n$ times, $freq(u,v)$ is equal to the proportion of the cases where the edge $(u,v)$ was selected. Note that each edge is here also directed. Besides, each edge does not receive the same amount of capacity: some edges have potentially more interfering edges and will be chosen less frequently. However, fairness among edges is respected.

The remaining description of the upper bound remains unchanged. The linear program {\sc lp} \ref{lp:vertex_upper_bound} keeps on holding, with the new values of $freq(u,v)$.

\section{Evaluation functions}
\label{section:evaluation}

We defined in the previous section a method to model the radio medium utilization. The radio resource sharing was translated in local constraints, in the form of inequalities. Therefore, any topology can conduct to a list of inequalities exhibiting the radio resource sharing. 
The objective of the work is to model the network capacity. Consequently, functions to evaluate such a capacity must be proposed. We propose here two objective functions: the first deals with the classical definition of capacity and the second function introduces the fairness.

\subsubsection{Max-Sum function}
The capacity is often described as the maximal throughput achievable in the network. Thus the objective function can be:
\begin{equation}
	Max\left( \sum_{p \in\mathcal{P}} f(p) \right)
\end{equation}
In such an approach, the network is considered in its totality: the objective is not individual. Thus, a network privileges some flows which present the lowest constraints on the radio interferences. In consequence, some flows will surely be null, because they consume too resources. Particularly, a flow can be multihops, meaning that a packet of this flow must be forwarded, and create more interferences. Consequently, the traffic pattern will surely be constituted by an independent set of single hop flows. Therefore, a determined source and destination couple can be impossible. According to us, this could constitute a misinterpretation of the real capacity of an ad hoc network for many applications. Nevertheless, this formulation gives an upper bound of the global achievable capacity.

\subsubsection{Max-Min function}
A fairness among the flows can be introduced. In such a case, the objective function can be formulated as:
\begin{equation}
	Max\left( Min_{p \in\mathcal{P}} f(p) \right)
\end{equation}
In such an approach, a minimal bandwidth is guaranteed for each flow in the network in such a way that the multihops flows are not handicapped. The global achievable throughput will surely be inferior to the \emph{max-sum} function case, since more interfering flows must cohabit. Nevertheless, we assume such an approach more representative of the general expected behavior of an ad hoc network. 
Additionally, the fairness ratio introduced in \cite{kumar05} could very well be adapted in this approach, although we chose to not give the corresponding results here since it represents more a \emph{tuning} parameter.

\section{Special-Case Study: the line}
\label{section:example}
We propose in this section to study the cases of the line to illustrate our proposition. The capacity is evaluated through the two objective functions max-sum and max-min. To simplify the explications, the control traffic is considered null. In the same way, the capacity is equal to one unit. In the \emph{max-min} case, $x$ represents the throughput of each node.

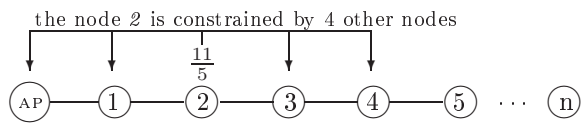
\begin{figure}[htb]
   \centering
      \begin{picture}(220,50)(0,0)
      	\put(3,3){\circle{15}}
      	\put(-1.5,1){\footnotesize{\sc ap}}
      	
      	\put(11,3){\line(1,0){18}}
      	
      	\put(35,3){\circle{12}}
      	\put(32,0){1}
      	\put(63,15){$\frac{11}{5}$}
      	\put(68,25){\line(0,1){5}}
      	\put(03,30){\line(1,0){129}}
      	\put(03,30){\vector(0,-1){15}}
      	\put(34,30){\vector(0,-1){15}}
      	\put(101,30){\vector(0,-1){15}}
      	\put(132,30){\vector(0,-1){15}}      	
      	
      	\put(5,32){\footnotesize the node \emph{2} is constrained by 4 other nodes}

      	\put(41,3){\line(1,0){20}}
      	
      	\put(68,3){\circle{12}}
      	\put(66,0){2}
      	
      	\put(75,3){\line(1,0){20}}
      	
      	\put(101,3){\circle{12}}
      	\put(99,0){3}
      	
      	\put(107,3){\line(1,0){20}}
      	
      	\put(133,3){\circle{12}}
      	\put(130,0){4}      	
      	
      	\put(139,3){\line(1,0){20}}
      	
      	\put(166,3){\circle{12}}
      	\put(163,0){5}    
      	
      	\put(180,2){\dots}  	
            	
      	\put(205,3){\circle{12}}
      	\put(203,0){n}      	
		\end{picture}
   \caption{the case of the line (pessimistic model)}
   \label{fig:line_case}
\end{figure}

Let be the topology illustrated in figure \ref{fig:line_case}. The Access Point is placed on the left extremity. The line contains $n$ nodes, and the {\sc ap}.

%
%

\subsection{Vertex-oriented fairness}

\begin{itemize}
	\item \emph{max-sum}: the bottleneck will appear in the node $1$. $1$ will forward all its traffic to the {\sc ap}. However, $1$ has four 2-neighbors. Thus, $1$ has $\frac{1}{5}$ of the medium capacity. The half of this capacity is reserved for the edge (1,{\sc AP}), and the other part for the edge (1,2). Finally, $\hbox{\emph{max-sum}}=\frac{1}{10}$.
	
		\item \emph{max-min}: the node $1$ will represent the bottleneck in the network. Let $x$ be the traffic sent by each node. $1$ must receive and forward the traffic of the $(n-1)X$ other nodes, and send its own traffic to the {\sc ap}. Since the node $2$ has four 2-neighbors and the node $1$ is a 2-neighbor of $2$, the capacity $\frac{1}{4+1}$ is allocated to the node $1$, which splits this capacity in the edge ({\sc ap}, 1) and (1, {\sc ap}). Finally, the capacity allocated to the node $1$ is used for the traffic of $n$ nodes: $\frac{1}{10} \leq (n-1)x + x$. Finally, $\hbox{\emph{max-min}}=\frac{1}{10n}$
\end{itemize}

\paragraph{Optimistic model}

\begin{itemize}
	\item \emph{max-sum}: the node $1$ sends all its traffic to the {\sc ap}. If the center is the node $1$, all its traffic and the traffic sent through (2, 3)\footnote{The bandwidth is shared among a \emph{center} and all its 2-neighbors} must be inferior to the capacity. In the same way, if $2$ is the center, the following stables exist:
	\begin{itemize}
			\item ({\sc ap}, 1) \& (3, 4)
			\item ({\sc ap}, 1) \& (4, 3)
			\item (1, {\sc ap}) \& (3, 4)
			\item (1, {\sc ap}) \& (4, 3)
	\end{itemize}
	Thus, $freq(1,${\sc ap}$)=\frac{1}{2}$. Since, the node $2$ sends no traffic, the node $1$ will have the total capacity. Finally, $\hbox{\emph{max-sum}}=\frac{1}{2}$ 

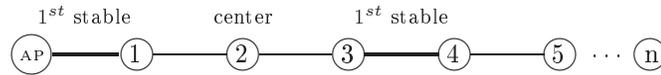
\begin{figure}[htb]
   \centering
      \begin{picture}(200,50)(0,0)
      	\put(3,3){\circle{15}}
      	\put(-1.5,1){\footnotesize{\sc ap}}
      	
      	\thicklines
      	\put(11,3){\line(1,0){25}}
      	\put(5,15){\footnotesize 1$^{st}$ stable}
      	\thinlines
      	
      	\put(43,3){\circle{12}}
      	\put(40,0){1}
      	
      	\put(49,3){\line(1,0){28}}
      	
      	\put(83,3){\circle{12}}
      	\put(80,0){2}
      	\put(72,15){\footnotesize center}
      	
      	\put(86,3){ \line(1,0){27}}
      	
      	\put(123,3){\circle{12}}
      	\put(120,0){3}
      	
      	\thicklines
      	\put(129,3){\line(1,0){28}}
      	\put(125,15){\footnotesize 1$^{st}$ stable}
      	\thinlines
      	
      	\put(163,3){\circle{12}}
      	\put(160,0){4}      	
       	
      	\put(169,3){\line(1,0){28}}
      	
      	\put(203,3){\circle{12}}
      	\put(200,0){5}      	
      	
      	\put(215,2){\dots}  	
      	
      	\put(237,3){\circle{12}}
      	\put(235,0){n}      	
    \end{picture}
   \caption{The case of the line (optimistic model)}
\end{figure}

	\item \emph{max-min}: the node $1$ will constitute here also the bottleneck. It sends $x$ traffic and must forward the traffic of $(n-1)$ nodes. The most-constrained set is the 2-neighborhood of the node $2$. The following holds: 
		\begin{itemize}
			\item Through the link $(1,\hbox{\sc ap})$ must be sent the traffic $x(n)$
			\item Through the link $(4,3)$ must be sent the traffic $x(n-3)$
			\item The node $2$ must additionally send the traffic $x(n-1)$
		\end{itemize}
		The frequencies computed for the \emph{max-sum} objective do not change. Thus, we have the following constraints:
		\begin{equation}
			x(n) 	 \leq	\left[ 1 - x(n-1) \right] \cdot \frac{1}{2} \footnotemark{}
		\end{equation}
		\footnotetext{Since the center is $2$, the traffic allocated to $(1,\hbox{\sc ap})$ must be inferior to the bandwidth minus the traffic of $2$ multiplied by the frequency of the link $(1,\hbox{\sc ap})$}
		\begin{equation}
			x(n-3)  \leq	\left[ 1 - x(n-1) \right] \cdot \frac{1}{2} \footnotemark{}
		\end{equation}
		\footnotetext{Since the center is $2$, the traffic allocated to $(4,3)$ must be inferior to the bandwidth minus the traffic of $2$ multiplied by the frequency of the link $(4,3)$}

		Finally, $\hbox{\emph{max-min}}=\frac{1}{3n-1}$ 
\end{itemize}

%
%

\subsection{Edge-oriented fairness}

\paragraph{Pessimistic model}
The node $1$ keeps on constituting the bottleneck. The 2-neighborhood of the edge $(2,3)$ constitutes the highest constraint. The capacity $\frac{1}{10}$ is allocated to each edge of this set since exactly 10 edges are 2-neighbors of $(2,3)$.

\begin{itemize}
	\item \emph{max-sum}: The node $1$ sends all its traffic to the {\sc ap} via the link $(\hbox{\sc ap},1)$. Thus, $\hbox{\emph{max-sum}}=\frac{1}{10}$

	\item \emph{max-min}: The node $1$ must send its own traffic and forward the traffic of the line, $n-1$ nodes. In consequence, $\hbox{\emph{max-min}}=\frac{1}{10n}$. 

\end{itemize}

We can remark that for the line, the \emph{max-min} remains unchanged whatever the fairness is (vertex-oriented or edge-oriented).

\paragraph{Optimistic model}
Since the frequencies do not change, the upper bound remains unchanged for the edge oriented fairness. 

\section{Results}
\label{section:results}

\subsection{Input Parameters}
Our objective is to estimate the capacity inherent to different routing protocols. To reach this goal, the behavior of some routing protocols were simulated with OPNET Modeler \cite{opnet}. More precisely, topologies from 20 to 60 nodes with an average degree of 20 nodes were generated. Then, OLSR, Wu \& Li, \somom and VSR were implemented and simulated on these topologies. The computed routes and the overhead were directly extracted from the simulations to be used in our {\sc lp} formulation. We chose to model 2 different traffic patterns: in an ad-hoc network, all the possible routes are computed, in an hybrid network, only routes toward the Access Point are computed.

To have a representative view of the different routing approaches, 3 main protocols were simulated:
\begin{itemize}
	\item OLSR: this protocol is relevant to represent the behavior of flat routing protocols. Particularly, shortest routes are computed, and the overhead is extracted directly from the simulations. Other routing protocols computing the same routes will surely offer a similar capacity.
	
	\item Wu \& Li - Routing via the backbone: Only the backbone topology is used, radio links between two backbone clients being considered logically deactivated. It is important to notice that in this case, the useless radio links keep on creating radio interferences, decreasing potentially the capacity. Besides, the original algorithm of Wu \& Li \cite{wu99} proposes to compute a backbone such that the shortest routes pass through the backbone. However, in this article, we used the most efficient algorithm \cite{carle04}: it optimizes the backbone cardinality, reducing the backbone redundancy. However, in this case, the backbone routes are not shortest routes. The overhead generated by a link-state routing protocol would be in this case reduced but the route length can increase.
		
	\item VSR \& \somom: In the ad hoc approach (VSR version), routes use the cluster topology, potentially longer than the shortest radio routes. In the hybrid approach (\somom version), routes use uniquely the backbone topology, for which the AP represents the root. Finally, the overhead is directly extracted from the simulations.
\end{itemize}

Finally, the following scheme was proposed to compute the network capacity:
\begin{enumerate}
	\item One topology of $x$ nodes was simulated ($x \in [20..60])$
	\item A routing protocol  was simulated giving overheads and routes (either in ad-hoc or in hybrid mode)
	\item Starting from the topology, the constraints modeling the radio resource sharing were extracted
	\item The constraints modeling the flows were obtained from the routes
	\item The capacity is finally computed with Cplex \cite{cplex} from this list of constraints and the objective function (either $max-sum$ or $max-min$ function, cf. section \ref{section:evaluation}). 
\end{enumerate}

\subsection{Results}
In this section, we investigate the capacity of the flat and self-organized structured routing protocols. We assume that the radio bandwidth is normalized to 1. First, general remarks on the evolution of the capacity with an increasing number of nodes is given. Then, the capacities inherent to different routing protocols in an ad-hoc and hybrid network are compared.

\subsubsection{General evolution of the capacity}
\begin{figure}
	\centering
		\includegraphics[width=8.5cm]{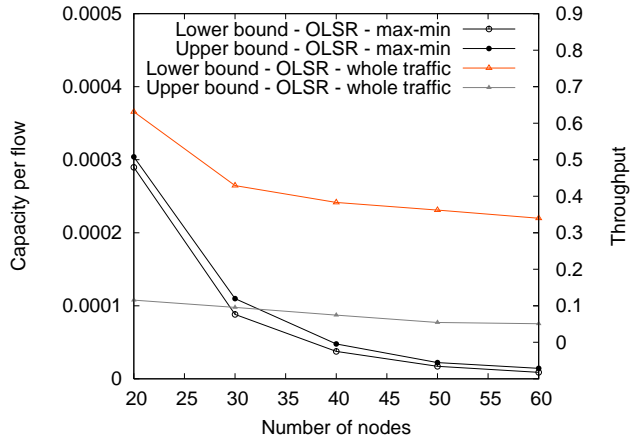}
	\caption{Comparison of the objective functions in an ad-hoc network using flat routing (link-oriented fairness)}
	\label{fig:objective_functions}
\end{figure}

First, the general evolution of the capacity in MANET is evaluated with the link-oriented fairness, maximizing the minimum capacity allocated to each flow in a flat network (fig. \ref{fig:objective_functions}). In a MANET, all the possible routes are computed. Thus, if the network comprises $n$ nodes, $n(n-1)$ paths are computed. With a flat routing protocol, we can remark that the capacity per flow decreases when the number of nodes increases: the number of flows grows, creating potentially more contentions. Consequently, the bandwidth allocated to each flow will surely decrease, corroborating the results of \cite{gupta00}. Oppositely, the total aggregated throughput sent across the network remains constant: many flows will with high probability pass through the center of the network since shortest paths are used. Thus, the center will represent a bottleneck, limiting the spatial reutilization of frequencies. Finally, we can note that the optimistic and the pessimistic resource sharing present a very close capacity.

\subsubsection{Ad-hoc networks}

\begin{figure}
	\centering
		\includegraphics[width=8.5cm]{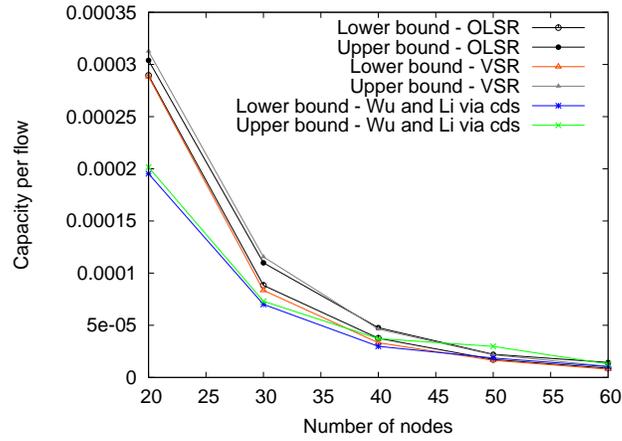}
	\caption{Capacity of an ad-hoc network with the max-min objective (link-oriented fairness)}
	\label{fig:max_min_ad_hoc_edge}
\end{figure}

The capacity of ad-hoc networks according to different routing protocols is evaluated. In a first time, the capacity with the max-min objective and with a link-oriented fairness bandwidth sharing is studied (fig. \ref{fig:max_min_ad_hoc_edge}). We can remark for the same reason as described previously that the capacity decreases when the number of nodes increases. The flat routing protocol presents the highest capacity: shortest routes in the initial network limit the route length, and consequently present less interferences. The VSR protocol which uses the backbone only for flooding, and computes shortest routes on the cluster topology does not increase greatly the length. Thus, the capacity of VSR and a flat routing protocol are very close in MANET. Oppositely, Wu \& Li which computes shortest routes through the backbone tends to lengthen the average route. Besides, only backbone nodes will route packets, creating a bottleneck in the network. In consequence, Wu \& Li protocol presents the lowest capacity.

\begin{figure}
	\centering
		\includegraphics[width=8.5cm]{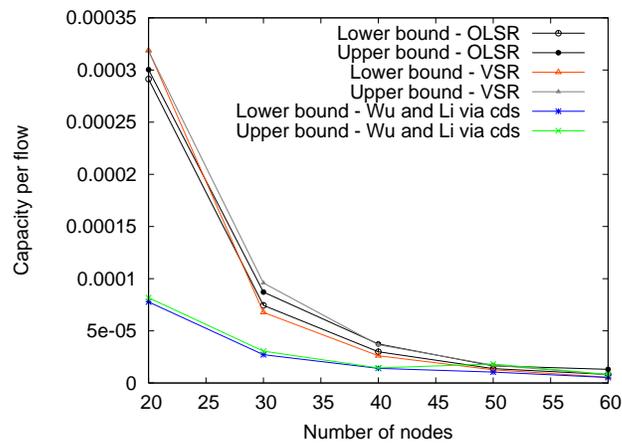}
	\caption{Capacity of an ad-hoc network with the max-min objective (node-oriented fairness)}
	\label{fig:max_min_ad_hoc_vertex}
\end{figure}

In a second time, the capacity is evaluated with the node-oriented fairness bandwidth sharing (fig. \ref{fig:max_min_ad_hoc_vertex}). The capacity of OLSR and VSR is very slightly impacted by a different fairness model. On the other hand, the capacity presented by Wu \& Li is lower: the backbone nodes concentrate all the traffic. Thus, the backbone clients receive a disproportioned bandwidth, creating oppositely a bottleneck in the backbone.

\begin{figure}
	\centering
		\includegraphics[width=8.5cm]{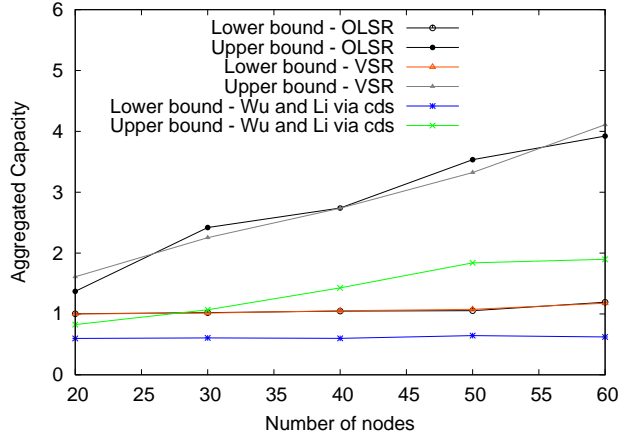}
	\caption{Capacity of an ad-hoc network with the max-sum objective (link-oriented fairness)}
	\label{fig:max_sum_ad_hoc_edge}
\end{figure}

Finally, we maximize the global network throughput using the \emph{max-sum} objective (fig. \ref{fig:max_sum_ad_hoc_edge}). With such a maximization, short routes will be advantaged since they create less radio interferences. Thus, the global capacity does not decrease when the network cardinality increases. We can even remark that with the optimistic bandwidth sharing, the global capacity increases: the degree being constant, the diameter increases, allowing a spatial reutilization of the radio medium. This corroborates the results of \cite{gupta00}. Oppositely, the pessimistic resource sharing tends to over-estimate the interferences, limiting the spatial re-utilization in small networks. 
Besides, we can remark that OLSR and VSR present a very close capacity, whatever the objective function is. The capacity of Wu \& Li remains much lower. Furthermore, the capacity in an optimistic resource sharing does not increase as fast as in OLSR or VSR. In conclusion, in an ad hoc network, a self-organization scheme does not impact severely on the capacity: OLSR and VSR offer a very similar capacity. Although routes are sometimes longer, and some nodes are overloaded, the network does not exhibit severe bottlenecks.

\subsubsection{Hybrid networks}

\begin{figure}
	\centering
		\includegraphics[width=8.5cm]{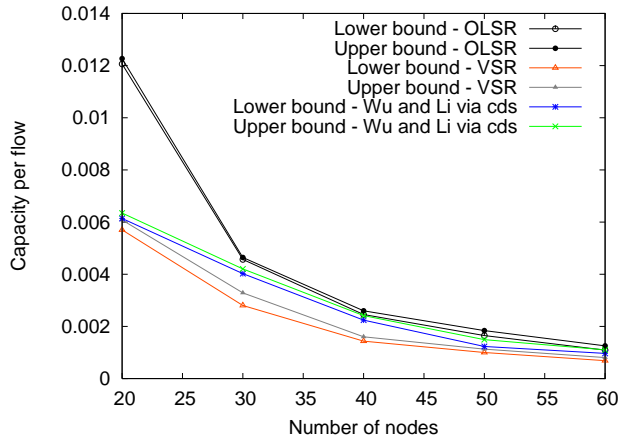}
	\caption{Capacity of an hybrid network with the max-min objective (link-oriented fairness)}
	\label{fig:max_min_hybrid_edge}
\end{figure}

\begin{figure}
	\centering
		\includegraphics[width=8.5cm]{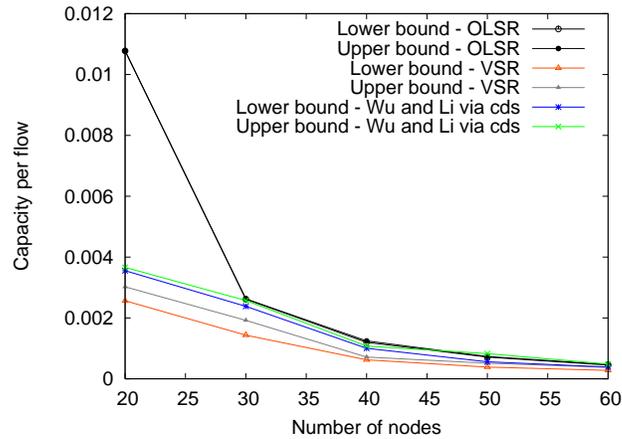}
	\caption{Capacity of an hybrid network with the max-min objective (node-oriented fairness)}
	\label{fig:max_min_hybrid_vertex}
\end{figure}

In a second time, the capacity with the max-min objective and the link-oriented fairness sharing in hybrid networks is studied (fig. \ref{fig:max_min_hybrid_edge}). In an hybrid network, the Access Point constitutes either the source or the destination of each flow. Thus, in a network with $n$ nodes, exactly $n$ flows exist. More precisely, since each flow creates a bidirectional traffic, $2n$ routes are computed. Consequently, the capacity per flow is much higher than in a MANET (fig. \ref{fig:max_min_hybrid_edge} and \ref{fig:max_min_hybrid_vertex}). OLSR offers an higher capacity than \somom and Wu \& Li: the Access Point represents the bottleneck of the hybrid network, but the flat routing protocol distributes efficiently the route. The backbone of Wu \& Li presents an higher throughput than \somom: the first one seems more efficient in hybrid networks to distribute the load among the backbone nodes. 

Finally, we can remark that the fairness model does not present any impact in an hybrid network: the capacity with the node-oriented fairness resource sharing (fig. \ref{fig:max_min_hybrid_vertex}) remains unchanged. 

In hybrid networks, a self-organization protocol seems to offer a degraded capacity compared to a flat routing protocol. Thus, an efficient backbone construction protocol optimizing the load distribution among the neighbors of the AP must be proposed. 

\section{Conclusion and future work}
\label{section:conclusion}
This paper focuses on generic methods for evaluating the capacity of MANets. Our approach consists in modeling radio resource sharing principles of 802.11-like MAC protocols as a set of linear constraints. We propose two MAC layer fairness models. One assumes that the probability to access the radio channel is uniformly distributed among the nodes, while the other one assumes that for the radio links. For each of these fairness models, we propose a pessimistic and an optimistic scenario of the spatial-reutilization of the medium. 

These models are not integrated with a particular routing scheme. They just need as input the overhead and routes for a given routing protocol. This work constitutes therefore a relevant framework for evaluating and comparing the capacity of different routing protocols.

We apply this framework to achieve a comparative analysis of flat and self-organized structured routing protocols. We have shown that the capacity provided by VSR and a flat routing protocol (OLSR) are similar.  Self-organization schemes seem therefore to be a relevant choice for routing in MANets, since they combine robustness, scalability and efficiency.  However, the strategy proposed by Wu \& Li, which consists in sending all the traffic through the backbone, has a too strong impact on the capacity to be practical.

When considering hybrid networks, both self-organized structures follow the Wu \& Li strategy, consequently providing a much lower capacity than the flat routing protocol. Self-organization schemes must therefore be improved for becoming an eligible solution. A relevant strategy seems to try to distribute the traffic in the neighborhood of the access point, while current schemes concentrate it on a few radio links.

In the close future, we plan to pursue our comparison campaign by including the other main flat routing protocols (AODV, DSR ,
DSDV\dots), even though we conjecture that there performances should theoretically be similar to those of OLSR. We are also interested in evaluating multi-path routing protocols \cite{bredin05}, since they have been proposed for improving the throughput of the network.

This work yields also an appealing question. Given an estimation of the capacity of a network, how to design a distributed routing protocol or a self-organization scheme in order to provide the maximal throughput ? Besides, we are looking for more realistic models for the (un-)fairness of 802.11 MAC layer, and willing to investigate its impact on the overall capacity of MANets .

\bibliography{rr_inria_capa}

\end{document}